%% file: main.tex
\documentclass[sn-mathphys-num 
              ]{sn-jnl}
\usepackage{graphicx}%
\usepackage{multirow}%
\usepackage{amsmath,amssymb,amsfonts}%
\usepackage{amsthm}%
\usepackage{mathrsfs}%
\usepackage[title]{appendix}%
\usepackage{xcolor}%
\usepackage{xspace}%
\usepackage{textcomp}%
\usepackage{manyfoot}%
\usepackage{booktabs}%
\usepackage{algorithm}%
\usepackage{algorithmicx}%
\usepackage{algpseudocode}%
\usepackage{listings}%
\usepackage{lineno}%
\usepackage{rotating}%

\usepackage{siunitx}

\usepackage{newfloat}
\DeclareFloatingEnvironment[
  name={Extended Data Figure},
  fileext=lox
]{extfigure}

\newcounter{exttable}
\newenvironment{exttable}[1][htbp]{%
    \begin{table}[#1]%
    \setcounter{table}{\value{exttable}}%
    \renewcommand{\tablename}{Extended Data Table}%
}{%
    \global\setcounter{exttable}{\value{table}}%
    \end{table}%
}

\newcommand{\AdamParke}{ALP\xspace}
\newcommand{\AlexanderKuhl}{AK\xspace}
\newcommand{\AnneAmyKlein}{AAK\xspace}
\newcommand{\BenjaminPointard}{BP\xspace}
\newcommand{\BiplabDutta}{BD\xspace}
\newcommand{\CeciliaClivati}{C.~Clivati\xspace}
\newcommand{\ChenHaoFeng}{CHF\xspace}
\newcommand{\ChetanVishwakarma}{CV\xspace}
\newcommand{\ChristianChardonnet}{C.~Chardonnet\xspace}

\newcommand{\DavideCalonico}{DC\xspace}
\newcommand{\ErikBenkler}{EB\xspace}
\newcommand{\EtienneCantin}{EC\xspace}
\newcommand{\FilippoLevi}{FL\xspace}
\newcommand{\HeinerDenker}{HD\xspace}

\newcommand{\IanHill}{IRH\xspace}
\newcommand{\IchiroUshijima}{IU\xspace}
\newcommand{\IngoNosske}{IN\xspace}
\newcommand{\IreneGoti}{IG\xspace}
\newcommand{\JacquesOlivierGaudron}{JOG\xspace}
\newcommand{\JacobTunesi}{JT\xspace}
\newcommand{\JeromeLodewyck}{JL\xspace}
\newcommand{\JochenKronjaeger}{J.~Kronj\"ager\xspace}
\newcommand{\JoshuaKlose}{J.~Klose\xspace}
\newcommand{\KilianStahl}{KS\xspace}
\newcommand{\MarcoPizzocaro}{MP\xspace}
\newcommand{\MarcoSchioppo}{MS\xspace}
\newcommand{\MasaoTakamoto}{MT\xspace}
\newcommand{\MatthewJohnson}{MYHJ\xspace}
\newcommand{\OlivierLopez}{OL\xspace}
\newcommand{\PaulEricPottie}{PEP\xspace}
\newcommand{\RodolpheLeTargat}{RLT\xspace}
\newcommand{\ShamboMukherjee}{SM\xspace}
\newcommand{\SimoneDonadello}{S.~Donadello\xspace}
\newcommand{\SoerenDoerscher}{S.~D\"orscher\xspace}
\newcommand{\StefanoCondio}{SC\xspace}
\newcommand{\ThomasLegero}{T.~Legero\xspace}
\newcommand{\TimLuecke}{T.~L\"ucke\xspace}
\newcommand{\TommasoPetrucciani}{TP\xspace}
\newcommand{\UweSterr}{US\xspace}
\newcommand{\WeiHuang}{WH\xspace}
\newcommand{\XiZhang}{XZ\xspace}
\newcommand{\YogeshwarKale}{YBK\xspace}

\newcommand{\PTB}{1}
\newcommand{\NPL}{2}
\newcommand{\LUH}{3}
\newcommand{\LPL}{4}
\newcommand{\LNEOP}{5}
\newcommand{\TODAI}{6}
\newcommand{\RIKENQML}{7}
\newcommand{\RIKENCAP}{8}
\newcommand{\UoB}{9}
\newcommand{\INRIM}{10}

\title{
    Optical clock network tracks tidal variations of the gravity potential
}
\author[\PTB{}]{\fnm{Kilian} \sur{Stahl}}
\equalcont{These authors contributed equally to this work.}
\author[\PTB{}]{\fnm{Ingo} \sur{Nosske}}
\equalcont{These authors contributed equally to this work.}
\author[\PTB{}]{\fnm{Chetan} \sur{Vishwakarma}}
\author[\PTB{}]{\fnm{Tim} \sur{L\"ucke}}
\author[\PTB{}]{\fnm{Joshua} \sur{Klose}}
\author[\PTB{}]{\fnm{Erik} \sur{Benkler}}
\author[\PTB{}]{\fnm{Alexander} \sur{Kuhl}}
\author[\PTB{}]{\fnm{Shambo} \sur{Mukherjee}}
\author[\PTB{}]{\fnm{Thomas} \sur{Legero}}
\author[\PTB{}]{\fnm{Jochen} \sur{Kronj\"ager}}
\author[\PTB{}]{\fnm{Uwe} \sur{Sterr}}
\author[\PTB{}]{\fnm{Christian} \sur{Lisdat}}
\author*[\PTB{}]{\fnm{S\"oren} \sur{D\"orscher}}\email{soeren.doerscher@ptb.de}
\author[\NPL{}]{\fnm{Chen-Hao} \sur{Feng}}
\author[\NPL{}]{\fnm{Matthew Y.\ H.} \sur{Johnson}}
\author[\NPL{}]{\fnm{Jacques-Olivier} \sur{Gaudron}}
\author[\NPL{}]{\fnm{Wei} \sur{Huang}}
\author[\NPL{}]{\fnm{Jacob} \sur{Tunesi}}
\author[\NPL{}]{\fnm{Xi} \sur{Zhang}}
\author[\NPL{}]{\fnm{Marco} \sur{Schioppo}}
\author[\NPL{}]{\fnm{Adam L.} \sur{Parke}}
\author*[\NPL{}]{\fnm{Ian R.} \sur{Hill}}\email{ian.hill@npl.co.uk}
\author[\LUH{}]{\fnm{Heiner} \sur{Denker}}
\author[\LPL{}]{\fnm{Biplab} \sur{Dutta}}
\author[\LPL{}]{\fnm{Etienne} \sur{Cantin}}
\author[\LPL{}]{\fnm{Olivier} \sur{Lopez}}
\author[\LPL{}]{\fnm{Anne} \sur{Amy-Klein}}
\author[\LPL{}]{\fnm{Christian} \sur{Chardonnet}}
\author[\LNEOP{}]{\fnm{Benjamin} \sur{Pointard}}
\author[\LNEOP{}]{\fnm{Rodolphe} \sur{Le Targat}}
\author[\LNEOP{}]{\fnm{J\'er\^ome} \sur{Lodewyck}}
\author[\LNEOP{}]{\fnm{Paul-Eric} \sur{Pottie}}
\author[\TODAI{},\RIKENQML{}]{\fnm{Ichiro} \sur{Ushijima}}
\author[\RIKENQML{},\RIKENCAP{}]{\fnm{Masao} \sur{Takamoto}}
\author[\TODAI{},\RIKENQML{},\RIKENCAP{}]{\fnm{Hidetoshi} \sur{Katori}}
\author[\UoB{}]{\fnm{Yogeshwar B.} \sur{Kale}}
\author[\INRIM{}]{\fnm{Stefano} \sur{Condio}}
\author[\INRIM{}]{\fnm{Irene} \sur{Goti}}
\author[\INRIM{}]{\fnm{Tommaso} \sur{Petrucciani}}
\author[\INRIM{}]{\fnm{Cecilia} \sur{Clivati}}
\author[\INRIM{}]{\fnm{Simone} \sur{Donadello}}
\author[\INRIM{}]{\fnm{Filippo} \sur{Levi}}
\author[\INRIM{}]{\fnm{Davide} \sur{Calonico}}
\author[\INRIM{}]{\fnm{Marco} \sur{Pizzocaro}}

\affil[\PTB{}]{\orgname{Physikalisch-Technische Bundesanstalt}, \orgaddress{\street{Bundesallee 100}, \postcode{38116} \city{Braunschweig}, \country{Germany}}}
\affil[\NPL{}]{\orgname{National Physical Laboratory}, \orgaddress{\street{Hampton Road}, \city{Teddington} \postcode{TW11~0LW}, \country{United Kingdom}}}
\affil[\LUH{}]{\orgdiv{Institut für Erdmessung}, \orgname{Leibniz Universit\"at Hannover}, \orgaddress{\street{Schneiderberg 50}, \postcode{30167} \city{Hannover}, \country{Germany}}}
\affil[\LPL{}]{\orgdiv{Laboratoire de Physique des Lasers (LPL)}, \orgname{Universit\'e Sorbonne Paris Nord, CNRS}, \orgaddress{\street{99 Avenue Jean-Baptiste Cl\'ement}, \postcode{93430} \city{Villetaneuse}, \country{France}}}
\affil[\LNEOP{}]{\orgdiv{Laboratoire Temps Espace (LNE-OP), Observatoire de Paris}, \orgname{Universit\'e PSL, Sorbonne Universit\'e, Universit\'e de Lille, LNE, CNRS}, \orgaddress{\street{61 avenue de l’Observatoire}, \postcode{75014} \city{Paris}, \country{France}}}
\affil[\TODAI{}]{\orgdiv{Department of Applied Physics, Graduate School of Engineering}, \orgname{The University of Tokyo}, \orgaddress{\city{Bunkyo-ku}, \state{Tokyo} \postcode{113-8656}, \country{Japan}}}
\affil[\RIKENQML{}]{\orgdiv{Quantum Metrology Laboratory}, \orgname{RIKEN}, \orgaddress{\city{Wako}, \state{Saitama} \postcode{351-0198}, \country{Japan}}}
\affil[\RIKENCAP{}]{\orgdiv{Space-Time Engineering Research Team}, \orgname{RIKEN Center for Advanced Photonics}, \orgaddress{\city{Wako}, \state{Saitama} \postcode{351-0198}, \country{Japan}}}
\affil[\UoB{}]{\orgdiv{School of Physics and Astronomy}, \orgname{University of Birmingham}, \orgaddress{\city{Birmingham} \postcode{B15~2TT}, \country{United Kingdom}}}
\affil[\INRIM{}]{\orgname{Istituto Nazionale di Ricerca Metrologica (INRiM)}, \orgaddress{\street{Strada delle Cacce 91}, \postcode{10135} \city{Torino}, \country{Italy}}}
\abstract{%
    Gravity affects the passage of time, as described by the General Theory of Relativity \cite{ein16}.
    On Earth, the rate of a clock as seen by a fixed observer changes by about one part in \num{e16} per metre in height \cite{cho10}.
    Chronometric levelling exploits this effect to directly measure differences in the gravity potential by comparing distant clocks \cite{ver83, bje85}.
    It has been proposed as a complementary approach to established geodetic techniques \cite{den17} for linking height reference systems and monitoring the terrestrial gravity potential \cite{bon15c, tan21b, vin24a}.
    Experimental demonstrations \cite{lis16, tak16, gro18a, tak20, gro24, icon24} have to date focused primarily on static potential differences.
    Yet, the potentials of the Moon, the Sun and other celestial bodies as well as the Earth's response to their tidal forcing change clock rates by several parts in \num{e17} \cite{voi16}, providing a predictable test signal to demonstrate optical clock networks resolving small variations in relativistic time dilation.
    Here, we report chronometric observations of time-varying gravity potential differences using a network of optical clocks and phase-stabilized fibre links in Europe.
    Two distinct geographical configurations, with baselines of approximately \qty{800}{\kilo\metre} each, result in different combinations of diurnal and semidiurnal tidal components.
    One of the comparisons uses a transportable optical clock deployed as a remote sensor.
    The measured relativistic rate shifts track the tidal variations, resolving the predicted signal within one day of measurement and agreeing with model predictions.
    These results extend experimental chronometric levelling from static to dynamic potential differences and
    highlight the need to account for the time dependence of relativistic time dilation in frequency metrology.
    They also open the prospect for quantum clock networks with improved clock stability and availability to move beyond known tides and monitor less well known gravity potential variations, such as those arising from surface deformation and mass redistribution in geologically or volcanically active regions \cite{bon15c, tan21b}.
}
\begin{document}
%
\maketitle
%
%
%

\section*{Introduction}
The notion of describing the figure of the Earth in relation to an equipotential surface of its gravity field dates back about two hundred years to the works of Gauss \cite{gau28}.
Surveying Earth’s gravity potential remains essential in geosciences as the foundation of physical height reference systems.
Measurements of its temporal variations constrain models of mass redistribution among continental water, ice, oceans and atmosphere \cite{unr15, hof06, tap04}. 
Gravity potential measurements are particularly valuable for processes that are difficult to monitor by other means, such as groundwater storage or deep-ocean currents and aquifers \cite{tap04}.  
Satellite gravity missions provide global observations of broad-scale potential variations \cite{tap04}, whereas potential differences between fixed points are often determined by a combination of geometric levelling, height measurements based on Global Navigation Satellite Systems (GNSS), and local gravimetry \cite{hof06}. 

Chronometric levelling provides a complementary means to determine gravity potential differences through the resulting differences in time dilation between distant atomic clocks.
These are measured as relativistic rate shifts (RRSs) of clock frequency ratios in comparisons via suitable time or frequency links \cite{ver83, bje85, meh18, del19}.
Experiments have progressed from initial demonstrations \cite{cho10, lis16, tak16, gro18a} to measuring potential differences with state-of-the-art height uncertainties of few centimetres \cite{icon24}.
Temporal variations of the potential difference between two locations have not been resolved using optical clocks, owing to the precision required to resolve the signal at its characteristic time scale.
Challenges range from the frequency stability and availability of clocks and links to their reproducibility if slow variations are to be investigated.
Furthermore, the potential difference between two clocks depends on geographical configuration, e.g., the uniformity of the time-varying potential components may lead to a strong supression across short distances.

Tidal potential variations have been modelled and removed as a background signal in some clock comparisons \cite{del17, icon24, hou26, tof26}.
Yet, they also provide a known signal to test an optical clock network's ability to resolve geodynamics.
Their leading contributions result from the tide-generating potential of the Moon and the Sun as well as the solid Earth tides and ocean tidal loading \cite{voi16}. 
Here, the solid Earth tides represent the response of the Earth to the tide-generating potential, which consists of surface deformation and the resulting mass redistribution.
These two effects counteract and reinforce the effect of the tide-generating potential, respectively. 
Ocean tidal loading includes the effect of water displacement and the resulting effects like the deformation of the Earth’s surface.
Variations caused by other tidal and non-tidal effects at similar time scales are usually at least one order of magnitude smaller \cite{voi16}. 
Overall, the dominant components of the tidal potential variations have diurnal and semidiurnal periodicity and change the RRS at an individual fixed location on the Earth's surface by up to \num{5e-17}.
The differential potential variations between locations depend on the baseline geometry:
longitudinal separation results in phase shifts, while latitudinal separation produces amplitude differences.
Across continental distances, this results in peak-to-peak differential RRS variations of several \num{e-18} or more.
Resolving them thus requires relative comparison instabilities in the range of \num{1e-18} at a few hours of averaging time.

Here, we demonstrate first tracking of the time-varying gravity potential using comparisons of \textsuperscript{87}Sr and \textsuperscript{171}Yb optical lattice clocks across the European fibre link network, providing high-resolution, local probes into the gravity potential.
We compared the self-contained transportable optical clock PTB-Sr4, which was deployed to the Istituto Nazionale di Ricerca Metrologica (INRiM) in Torino, Italy, to the laboratory clock PTB-Sr3 at the Physikalisch-Technische Bundesanstalt (PTB) in Brunswick, Germany.
We further present a complementary measurement between the laboratory clocks NPL-Sr1 and PTB-Sr3, which are located at the National Physical Laboratory (NPL) in Teddington, United Kingdom and at PTB, respectively, as well as results from other clock comparisons.
We demonstrate that both NPL-Sr1 and PTB-Sr4 track and resolve the predicted tidal RRS variations in less than one day of measurement when compared to the laboratory clock PTB-Sr3.

\section*{Results}
\begin{figure}
    \centering
    \includegraphics[width=1\linewidth]{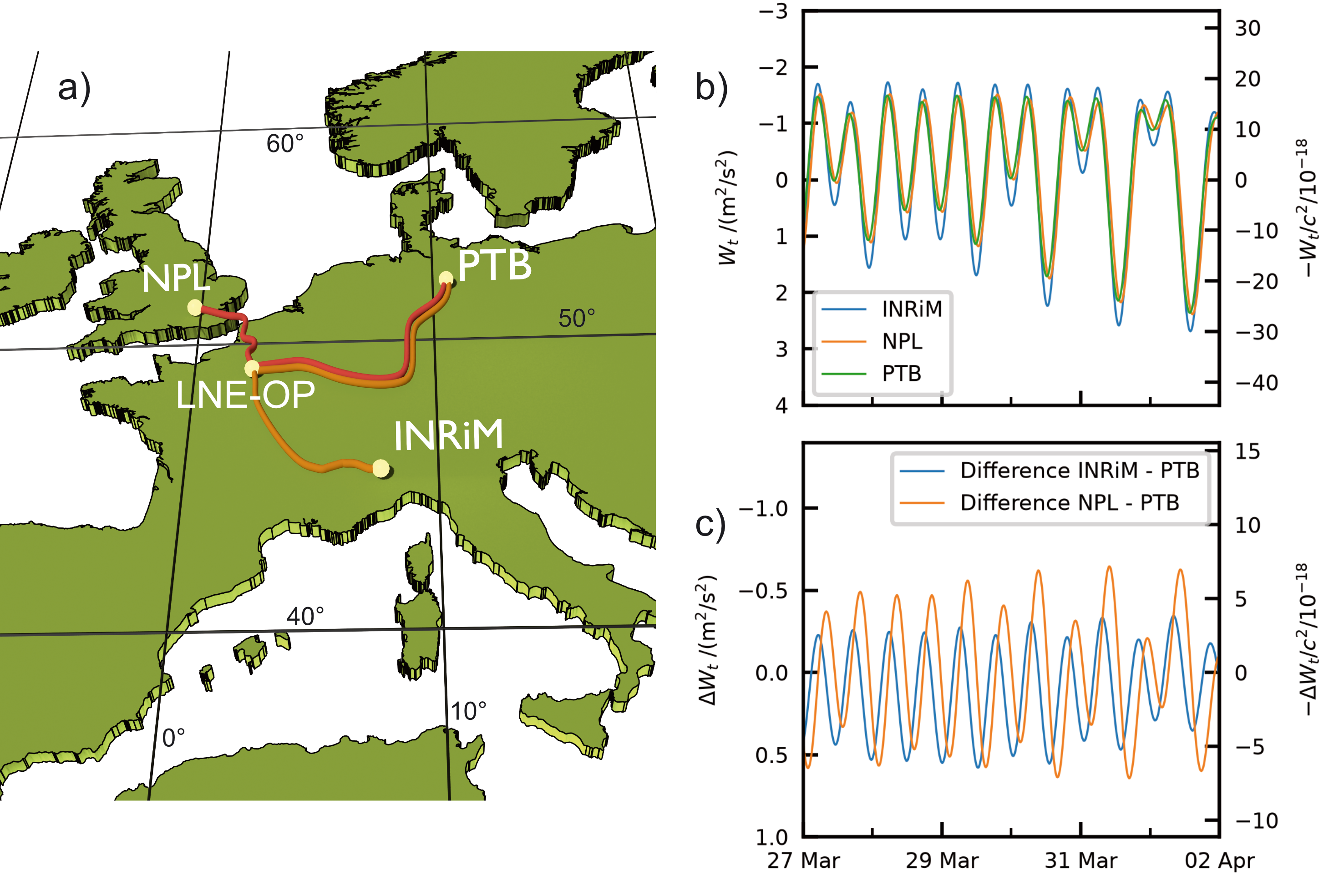}
    \caption{
        \textbf{a)}
        Geographical layout of the optical clock network.
        PTB is connected to INRiM (orange connection) and NPL (red connection) via a series of optical fibre links and frequency combs.
        Only parts of the network that are used in this work are shown.
        \textbf{b)}
        Deviation $W_\mathrm{t}$ of the instantaneous gravity potential from the static value due to tidal effects and corresponding RRS $-W_\mathrm{t}/c^2$ from time dilation for a short timespan in 2025.
        \textbf{c)}
        Differences of $W_\mathrm{t}$ between either INRiM or NPL and PTB.
    }
    \label{fig:map}
\end{figure}
Optical lattice clocks at four institutes are interconnected by the phase-stabilized interferometric fibre links of the European optical network \cite{lis16, sch22a, cli22b, lin25a, icon24, piz26}, with optical frequency combs connecting the clocks to the fibre links.
Figure~\ref{fig:map}a shows the geographical layout of this optical clock network.
The Laboratoire de Physique des Lasers and Laboratoire Temps Espace (LNE-OP) in Paris act as a central node.
We report on measurements carried out in three comparison campaigns in March 2023 (NPL--PTB), March 2025 (INRiM--PTB) and June 2025 (NPL/LNE-OP--PTB),
noting that the 2023 clock comparisons have been evaluated previously in a different context \cite{icon24}.
In this article, we focus on resolving tidal RRS variations between INRiM and PTB as well as between NPL and PTB.
Both comparisons cover geodetic distances of around \qty{800}{\kilo\meter} ($d_{\textrm{INRiM},\textrm{PTB}} \approx \qty{840}{\kilo\meter}$, $d_{\textrm{NPL},\textrm{PTB}} \approx \qty{740}{\kilo\meter}$).

The tide model that we use predicts variations of the RRS due to the instantaneous gravity potential by about \num{4e-17} at each station (Fig.~\ref{fig:map}b, see Methods on the tide model).
Since INRiM and PTB differ primarily in latitude ($\Delta\varphi \approx \qty{7}{\degree}$) and to a lesser extent in longitude ($\Delta\lambda \approx \qty{3}{\degree}$), the spectral constituents of the tidal RRS variations exhibit major differences in amplitude and minor shifts in phase.
In contrast, the RRS variations at NPL and PTB are almost exclusively phase-shifted as these stations are situated at similar latitudes ($\Delta\varphi \lesssim \qty{1}{\degree}$, $\Delta\lambda \approx \qty{11}{\degree}$).
The resulting differential RRS variations between the stations are up to \num{5e-18} from peak to peak (Fig.\ \ref{fig:map}c) and exhibit distinct spectral features:
Semidiurnal constituents are dominant in both cases,
but the variations between NPL and PTB further contain substantial diurnal constituents.

In order to track the oscillations, the clock comparisons should be able to resolve them within a single period.
The measurement instability caused by clock frequency noise, link phase noise and any other perturbations must therefore be smaller than the instability induced by the oscillation around its characteristic time scale.
In our case, this threshold corresponds to a fractional uncertainty at the low to mid-\num{e-18} level at averaging times $\tau$ around \qty{3}{\hour} (see Extended Data Figure~\ref{fig:ext:adev}).
The clock comparisons must reach a combined frequency instability $\sigma_y(\tau) \sim 5 \times 10^{-16}/\sqrt{\tau / \unit{\second}}$ or below, where $y$ is the reduced frequency ratio \cite{lod20} of the two clocks.
For remote clock comparisons, this remains a challenge and requires the use of high-performance ultrastable resonators, zero-dead-time clocks \cite{sch17}, or synchronous interrogation of both clocks \cite{tak11, tak16}, where the latter is hindered by short-term frequency transfer noise \cite{sch22a}.
Our measurements meet this requirement by leveraging the high stabilities of the laboratory ultrastable resonators at NPL \cite{sch22a} and PTB \cite{mat17a} as well as the transportable resonator of PTB-Sr4 \cite{her22}, reaching instabilities of $\sigma_y \approx 4.7 \times 10^{-16}/\sqrt{\tau / \unit{\second}}$ for the comparison between the transportable clock PTB-Sr4 deployed to INRiM and the laboratory clock PTB-Sr3 (INRiM--PTB) and $\sigma_y \approx 3.8 \times 10^{-16}/\sqrt{\tau / \unit{\second}}$ during the comparison between NPL-Sr1 and PTB-Sr3 in 2023 (NPL--PTB).

\begin{figure}
    \centering
    \includegraphics[width=1\linewidth]{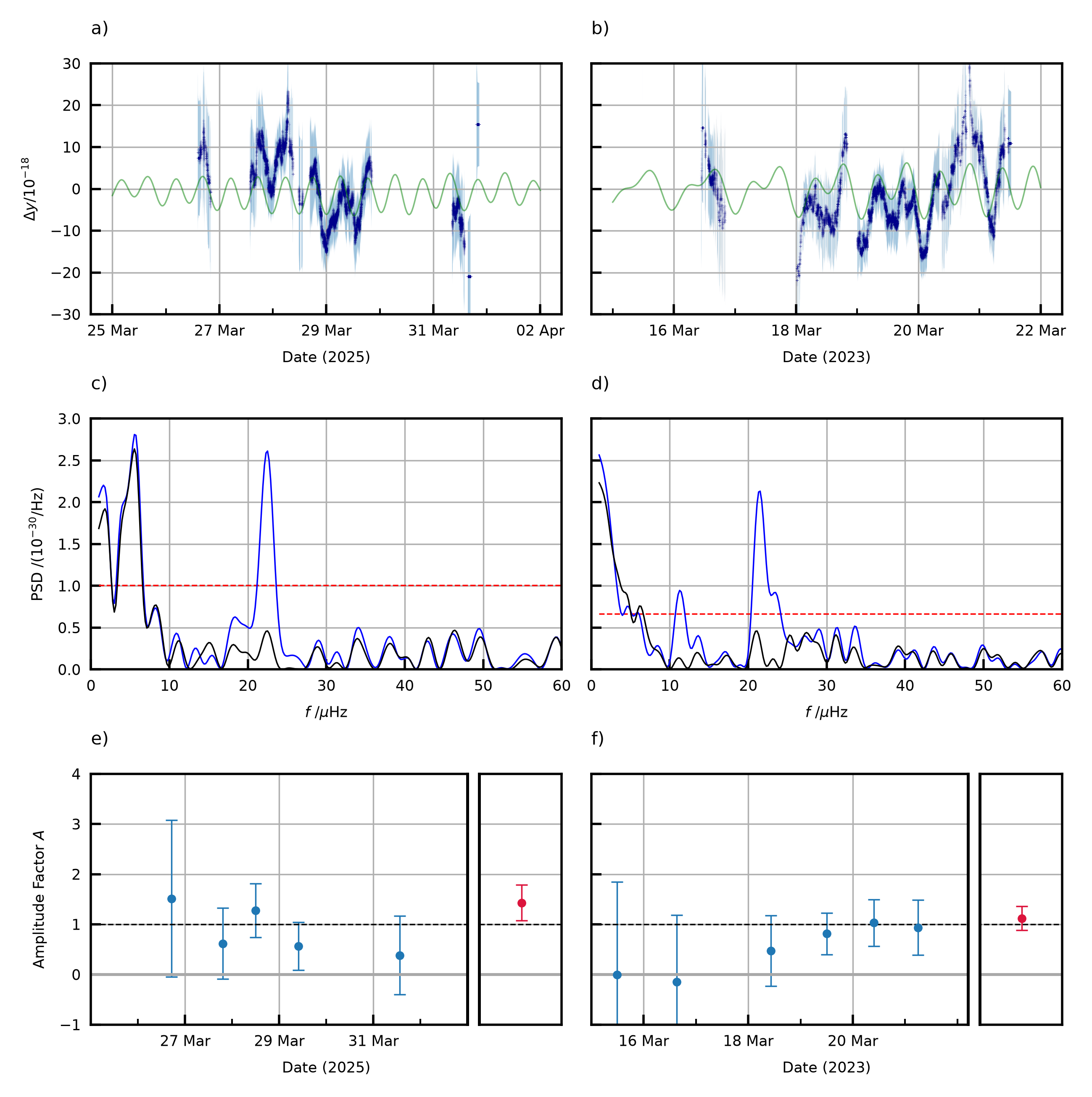}
    \caption{
        Overview of tide observations in the INRiM--PTB (left column) and the NPL--PTB (right column) clock comparisons.\\
        \textbf{(a--b)}:
        The solid green curve shows the predicted differential tidal RRS $\Delta W_\mathrm{t}/c^2$ as determined from the tide model.
        The blue curve shows a \qty{3}{\hour} moving average of the measured reduced frequency ratio's deviation from its mean, $\Delta y = y - \bar{y}$, where $\bar{y} = (-5.4, -33.6) \times 10^{-18}$, respectively.
        The light blue area indicates the estimated statistical uncertainty.\\
        \textbf{(c--d)}: 
        The blue curve shows the power spectral density (PSD) estimated as the periodogram of each measurement.
        The black curve is the PSD of the residuals after subtracting the predicted RRS variations.
        The dashed red line indicates the 99\textsuperscript{th} percentile of the white-frequency-noise PSD that we estimate from the instability of the measurement.\\
        \textbf{(e--f)}:
        The points show the results of the signal amplitude estimate of the Kalman filter.
        Daily values are shown in blue.
        The result of each whole dataset is shown in red in a separate field on the right side of each figure. 
        The expected value of the amplitude scaling factor, $A = 1$, is marked as a dashed black line.
    }
    \label{fig:freq-diff}
\end{figure}

In Figs.~\ref{fig:freq-diff}a and \ref{fig:freq-diff}b, we compare the measured reduced frequency ratios of the clocks to the RRS variations predicted by the tide model.
Here, we have low pass--filtered the observations to separate the tidal variations from noise by applying a \qty{3}{\hour} moving average,
which suppresses short-term clock and link noise effectively but retains the amplitude of the tidal variations \cite{tak16}.
As a result, we observe frequency ratio variations between the clocks that qualitatively resemble the predicted tidal RRS variations.

To underpin these observations quantitatively, we compute the Lomb--Scargle periodograms \cite{van18} (Figs.\ \ref{fig:freq-diff}c and \ref{fig:freq-diff}d) of the observed frequency ratios themselves as well as their residuals, i.e., after subtracting the predicted tidal RRS variations from the observations.
Here, taking the residuals before computing the periodograms carries information about the relative phases of the observed and predicted signal, even though the individual periodograms discard phase information.
As a measure of significance, we further indicate the 99\textsuperscript{th} percentile of periodogram power spectral density that arises from the white frequency noise floor estimated from each comparison's instability.
As expected, the frequency ratios exhibit significant semidiurnal oscillations ($f \approx \qty{23}{\micro\hertz}$), which are strongly suppressed in the residuals.
This confirms that the observed oscillations match the predictions by the tide model in both amplitude and phase.
The NPL--PTB comparison further exhibits significant diurnal ($f \approx \qty{12}{\micro\hertz}$) oscillations of the frequency ratio (Fig.\ \ref{fig:freq-diff}d), which similarly match the prediction in amplitude and phase.
By contrast, we neither expect nor observe substantial diurnal oscillations for the INRiM--PTB comparison (Fig.\ \ref{fig:freq-diff}c).

The Lomb--Scargle periodograms, however, do not readily quantify the time required to detect the tidal RRS variations.
Instead, we apply a progressive estimator to the measurements to investigate the time to detection as well as to demonstrate detection and tracking of arbitrary predicted signals.
The analysis is performed in back-testing here, but can be applied to real-time data without modification.
The estimator is a linear Kalman filter that tracks the tidal RRS variations by comparing the observed frequency ratio of the clocks to the instantaneous time-dependent RRS component predicted by the tide model.
It progressively estimates both the underlying frequency ratio of the clocks in the absence of the tidal variations and the observed variations' amplitude $A$ relative to the model prediction.
We account for possible unrelated variations of the frequency ratio (e.g., changing systematic biases), especially across gaps in the measurement time series, by incorporating non-zero process noise of the true frequency ratio, while treating the true amplitude as constant.
Here, the amplitude estimates also serve to quantify the estimator's detection performance, e.g., by considering the significance of their deviation from zero.
We apply the estimator to the full dataset of a comparison for estimating the overall detection confidence and to separate daily datasets for estimating the time to detection.
A detailed discussion is given in the Methods on Kalman filtering.

The amplitudes $A$ estimated by the Kalman filter for the INRiM--PTB and NPL--PTB comparisons are shown in Figs. \ref{fig:freq-diff}e and \ref{fig:freq-diff}f, respectively.
Daily estimates with sufficient uptime consistently agree with the expected value of one and deviate from zero by more than one standard deviation.
Overall, we find $A = 1.42(35)$ for the INRiM--PTB comparison and $A = 1.12(24)$ for the NPL--PTB comparison.
Both comparisons took place over the course of one week, including downtimes.
The results agree with the model prediction and provide highly significant detections, deviating from zero by more than four times their standard uncertainty.

\begin{figure}
    \centering
    \includegraphics[width=\linewidth]{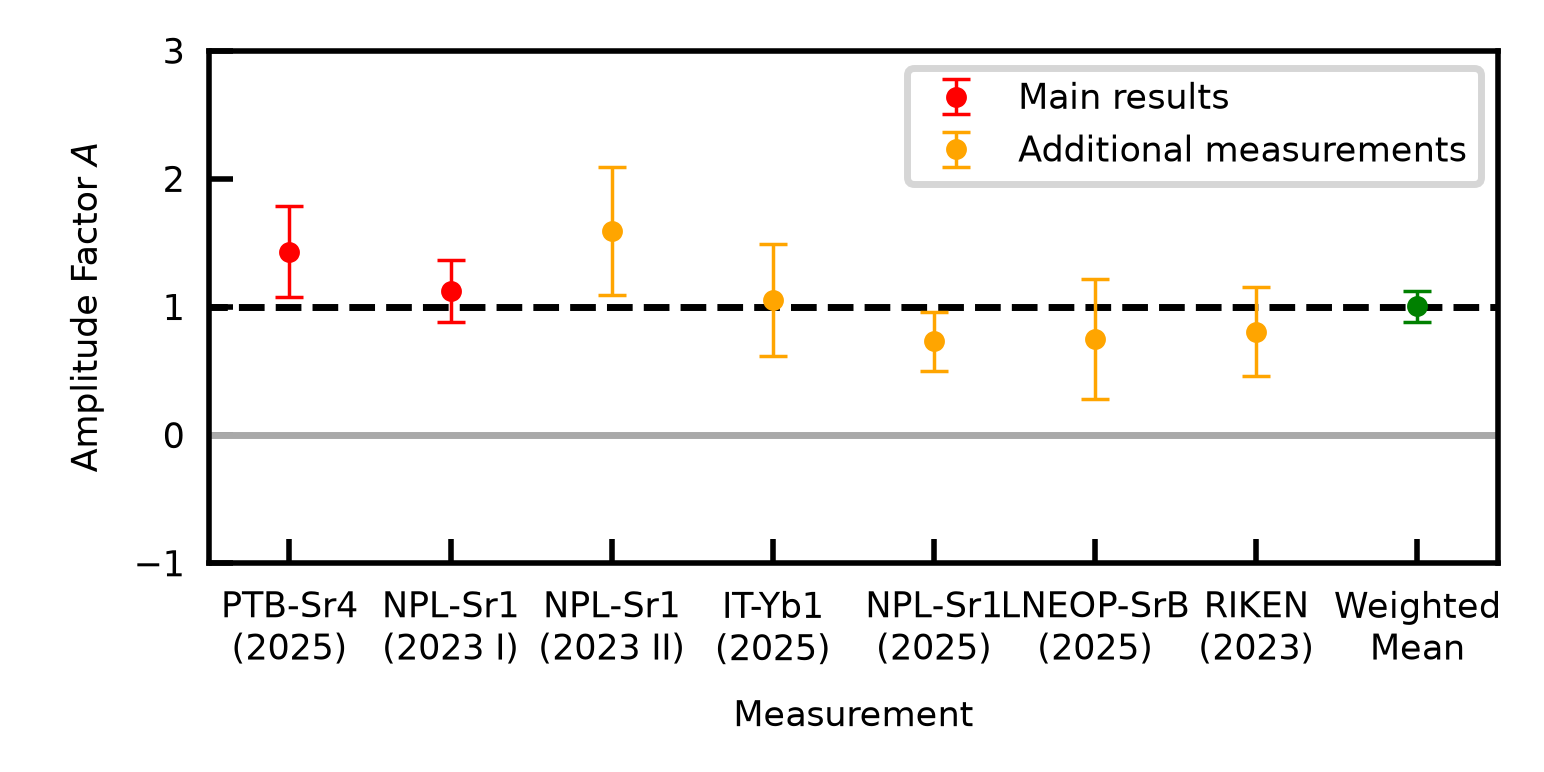}
    \caption{Results of the amplitude estimate of the tide signal by the Kalman filter for all clock comparisons reported in this work. Results discussed in the main text are highlighted in red.}
    \label{fig:kalman_total}
\end{figure}

The overall amplitude estimates of these and other comparisons of lattice clocks (with respect to the clock PTB-Sr3, see Methods and Extended Data Fig.~\ref{fig:ext:additional_clocks}) are shown in Fig.\ \ref{fig:kalman_total}.
The other comparisons often have a higher instability but still result in detection of the tidal RRS variations across a single measurement campaign.
The weighted mean amplitude over all measurements is $\bar{A} = 1.00(12)$, where we consider the individual results uncorrelated because the measurements are separate in time or not limited in instability by the common clock PTB-Sr3.
This result is in excellent agreement with the predictions of the tide model.

\section*{Discussion}
Previous experiments \cite{lis16, tak16, gro18a, tak20, icon24} have improved the static resolution of chronometric levelling to match the resolution of established geodetic methods \cite{den17}.
In this work, optical clock comparisons across baselines of about \qty{800}{\kilo\metre} have resolved time-varying tidal components of the gravity potential difference, which cause diurnal and semidiurnal modulations equivalent to changes on the order of \qty{10}{\centi\meter} in height at each station.
Our tide model accounts for the potential of the tide-generating celestial bodies as well as the solid Earth tides and ocean tidal loading that result from tidal forcing.
The observed variations of the clocks' frequency ratios reproduce the predicted tidal RRS variations well.
The baselines INRiM--PTB and NPL--PTB exhibit different amplitudes of diurnal and semidiurnal components due to their geographical configurations.
The corresponding spectral components of the frequency ratios are reduced to insignificant levels after subtracting the time-varying RRS components predicted by the tide model.
Moreover, the amplitude estimates from the Kalman filter differ from zero significantly and agree with the predicted amplitudes.
The agreement in the time domain, frequency domain and using the recursive estimator supports the interpretation of the observed rate shift variations as caused by the tidal RRS variations.

The optical clock network's ability to detect tidal RRS variations depends on the geographical configuration of the baseline and time of measurement as well as on the instability of the clock comparison, availability of the clock and fibre links and measurement duration.
During operation, the frequency-transfer links contributed to the instability at a level well below the contributions by the clocks at the relevant times $\tau$, and thus did not limit the measurement sensitivity.
Similar to other clock comparisons \cite{bel21, icon24, aep26}, we observed increased variability of the frequency ratio at long averaging times, departing from a white frequency noise--limited model.
Time-varying systematic biases are one possible cause of increased variability and not unexpected within the respective uncertainties;
they can be caused by changes in the operating conditions of a clock or by correcting time-varying systematic shifts statically, for instance.
However, no identified effect is expected to vary significantly at the phase and frequency of the predicted diurnal and semidiurnal tidal RRS oscillations.
In the present work, the increased variability does not prevent identification of the tidal signal, albeit it may limit the accuracy with which its amplitude can be determined.
Improvements of the stability and reproducibility of optical clocks (e.g., by deploying zero dead time clocks \cite{sch17} or utilizing quantum nondemolition measurements \cite{rob24}) will enable them to resolve small dynamic variations of gravity potential differences with even better resolution.

Transportable optical clocks have been used as travelling frequency standards for interlaboratory comparisons, static chronometric levelling and gravitational redshift tests \cite{gro18a, tak20, gro24, icon24}.
In this work, a transportable clock was deployed as a remote sensor of time-varying gravity potential differences.
It achieved a similar instability as the stationary systems, thanks to pre-stabilizing its clock laser to a transportable ultrastable optical cavity \cite{her22}.
This showcases the feasibility of deploying transportable clocks to stations that are selected for their expected geophysical signals rather than the availability of an established clock laboratory, provided that a suitable fibre link is available.

In the present study, the tides primarily provide a known time-dependent signal, allowing the clock comparisons to be tested against model predictions.
Conversely, tidal RRS variations also constitute a deterministic contribution to the instability and systematic bias of remote clock comparisons and need to be accounted for, even on long timescales.
For remote comparisons with similar instabilities as achieved here and below, the diurnal and semidiurnal components can become comparable to, or even exceed, the statistical noise at the relevant averaging times.
Modelling and removing the tidal RRS variations can reveal other low-frequency variations, including time-dependent systematic biases of the clocks, and improve the analysis of clock performance.
Notably, the mean RRS during a specific measurement interval can differ from the long-term average substantially, even for campaigns lasting several weeks or longer and achieving high availability.
This occurs in cases where the tidal RRS variations include significant constituents with periods around two weeks, one year, or even longer \cite{agn07}.
The introduced bias depends on the geographical configuration, measurement interval and temporal sampling, becoming relevant for remote clock comparisons targeting fractional uncertainties near \num{e-18} and below.

Monitoring the tidal RRS variations during operation of the optical clock network would facilitate handling their effect on measurements.
Kalman filtering provides a method for tracking time-varying signals during live measurements.
With suitable state and measurement models, the filter method could be adapted to other time-dependent signals (e.g., detection of clock frequency ratio oscillations due to differential exposure to dark matter \cite{rob20}) or to the detection of non-periodic perturbations (e.g., magma chambers filling \cite{bon15c}).

Accounting for the tides accurately is also relevant for applications of quantum clock networks in fundamental physics research, including tests of general relativity \cite{del17, tak20}, the detection of gravitational waves \cite{kol16, gan26} and the search for dark matter \cite{rob20, fil25}.
In searches for low-frequency signals (e.g., tests of special relativity \cite{del17}), the tidal RRS variations constitute a known, deterministic background that must be modelled and, where appropriate, removed.
Other experiments (e.g., gravitational redshift tests) require taking into account the difference between the long-term average of the RRS and its average over the measurement.

The expansion of fibre link networks would further broaden the set of accessible baseline geometries.
As the differential tidal signal is determined primarily by the differences in latitude and longitude,
tide signals can be larger across intercontinental baselines than across continental ones \cite{voi16}.
The semidiurnal component of the tidal RRS variations for stations at similar latitude is largest for a longitudinal separation of \qty{90}{\degree}, for instance.
Expanding fibre link networks could therefore allow the clock locations to be selected for optimal sensitivity to particular tidal components.
The European optical network employed in this work has connected Italy, France, Germany and the United Kingdom for some time \cite{lin25a}.
A recent expansion to Toru\'{n} in Poland has increased the maximum available longitudinal separation from \qty{11}{\degree} to approximately \qty{19}{\degree}.
Planned connections to Belgium and the Netherlands will provide a denser set of baseline geometries.
In the long term, transoceanic frequency transfer links \cite{mar22b} could provide longitudinal separations approaching those required for maximal sensitivity to the tides.
Simultaneous comparisons among multiple clocks over multiple baselines \cite{lin25a, piz26} would further add spatial information and help distinguish clock-local variations from spatially coherent changes in the gravity potential.

\section*{Summary}
In conclusion, we have combined laboratory optical clocks, a transportable optical clock, a continental-scale fibre link network and geodetic tide modelling to resolve time-varying gravity potential differences between distant stations, extending experimental chronometric levelling from static to dynamic signals.
Prospectively, improved availability and performance of laboratory and transportable optical clocks could establish chronometric levelling as a method complementary to GNSS, terrestrial gravimetry and satellite gravity missions.
These techniques observe different quantities:
GNSS constrains geometric displacement, terrestrial gravimetry measures local gravity acceleration, and satellite gravity missions (e.g., GRACE) measure large-scale gravity potential.
Chronometric levelling measures differences in gravity potential through the relativistic time dilation and allows measurements on a much finer mesh in both space and time.
Optical clock networks could therefore provide ground-based validation of satellite gravity missions and complement their monthly sampling with differential, daily to weekly observations \cite{sch21b}.
They can also constrain geodynamic parameters (e.g., Love numbers, which quantify the Earth's response to tidal forcing \cite{voi16}) and monitor potential changes associated with surface deformation or the redistribution of continental water, ice or atmospheric and oceanic masses \cite{bon15c, voi16, tan21b, sch21b}.

\section*{Methods}

\subsection*{Modelling the tidal RRS variations}
The time-varying component $y_\mathrm{t} = -W_\mathrm{t}/c^2$ of the RRS experienced by a clock results from the deviation of the instantaneous gravity potential from its long-term mean value, $W_\mathrm{t}(t) = W(t) - \left<W\right>$.

The mean potential of a clock is given by $\left<W\right> = W_\mathrm{ref} + \Delta W_\mathrm{ref \rightarrow ck} + \Delta W_\mathrm{MT}$, where $W_\mathrm{ref}$ is the zero-tide mean potential of a height reference marker at the respective station determined through the GNSS/geoid approach \cite{den17, rie20}.
We include a correction \cite{ihd08, icon24} $\Delta W_\mathrm{MT}$ to convert the potential from the zero-tide convention to the mean-tide convention \cite{mae21}, as the mean RRS between stations is given by the difference of the mean-tide potentials.
$\Delta W_\mathrm{ref \rightarrow ck} = g \, {\Delta H}$ is the potential difference between the clock and the reference marker, which is determined by combining the orthometric height difference $\Delta H$ measured by geometric levelling with the known local mean gravity acceleration $g$ \cite{rie20}.

The instantaneous gravity potential of a height reference marker, $W_\mathrm{ref}(t)$ is computed as described in a previous publication \cite{voi16}.
The model includes the tide-generating potential, the deformation of the solid Earth and induced potential changes, ocean tidal loading, solid Earth pole tides and length-of-day tides.
The stations' coordinates are $\varphi = \ang{45;00;55}$, $\lambda = \ang{7;38;20}$, $h = \qty{290}{\meter}$, $H = \qty{240}{\meter}$ for INRiM, $\varphi = \ang{51;25;30}$, $\lambda = -\ang{0;20;35}$, $h = \qty{55}{\meter}$, $H = \qty{10}{\meter}$ for NPL, $\varphi = \ang{48;50;7.5}$, $\lambda = -\ang{2;20;10}$, $h = \qty{100}{\meter}$, $H = \qty{10}{\meter}$ for LNE-OP, and $\varphi = \ang{52;17;45}$, $\lambda = \ang{10;27;35}$, $h = \qty{125}{\meter}$, $H = \qty{80}{\meter}$ for PTB, where $(\varphi, \lambda, h)$ and $H$ are the ellipsoidal coordinates and heights above sea level, respectively \cite{voi19}.
The time variation of the potential difference $\Delta W_{\mathrm{ref}\rightarrow\mathrm{clk}}$ can be neglected.

\subsection*{Optical clock network}
The European fibre link network for time and frequency metrology has been used in previous international comparisons of optical clocks \cite{lin25a, piz26, icon24}.
It connects institutes that operate optical clocks across Europe, including INRiM, LTE/LNE-OP, NPL and PTB.
In France and Italy, the network uses the national research infrastructure REFIMEVE \cite{can21} and the Italian Quantum Backbone national network \cite{cli20a}, respectively.
Optical frequency combs connect the ultrastable lasers of the optical clocks to the network at each institute.

For this work, we use optical lattice clocks at INRiM, LTE/LNE-OP, NPL and PTB.
The clocks are interconnected through the fibre links from INRiM (via Modane) \cite{cli22b}, NPL \cite{sch22a} and PTB (via Strasbourg) \cite{lis16} to Paris, France.
Transportable clocks were connected to the network via frequency combs operated by the respective host institutes.
The instability contributions from fibre links and combs are well below \num{1e-18} at averaging times of \qty{e3}{\second} \cite{piz26}.

The \textsuperscript{87}Sr or \textsuperscript{171}Yb optical lattice clocks NPL-Sr1 \cite{hob20a}, PTB-Sr3 \cite{klo26}, PTB-Sr4 \cite{nos25}, LNEOP-SrB  \cite{lod16} and IT-Yb1 \cite{got23} as well as the RIKEN Sr clock \cite{tak20} have been described in other publications.
Among these, PTB-Sr4 and the RIKEN clock are transportable clocks, which were deployed to INRiM in 2025 and NPL in 2023, respectively.
The systematic uncertainties of the clocks range from few \num{e-18} \cite{nos25} to about \num{1e-17} \cite{hob20a, icon24} for the two main comparisons but are of minor concern here as variations of the systematic biases typically remain small over the course of a measurement campaign.

The frequency instabilities of the lattice clocks are limited by the Dick effect \cite{dic87}, which is the aliasing of laser frequency noise due to the temporally varying sensitivity of atomic interrogation (e.g., due to dead times).
Therefore, clock instabilities in these measurements are governed mainly by their ultrastable optical reference resonators.
NPL-Sr1 uses a room-temperature, ultra-low expansion glass (ULE) cavity with a length of $\qty{48.5}{\cm}$ and a fractional frequency instability of down to $7 \times 10^{-17}$ \cite{sch22a} and reaches a fractional frequency instability $\sigma_y < \num{3e-16} / \sqrt{\tau / \unit{\second}}$ \cite{icon24}.
PTB-Sr3 uses an ultrastable cryogenic silicon resonator with a frequency instability $4 \times 10^{-17}$ \cite{mat17a} (with stability transfer via a frequency comb and a $\qty{48}{\centi\meter}$--long, room-temperature ULE cavity \cite{hae15a}) and reaches an instability $\sigma_y \lesssim \num{1e-16} / \sqrt{\tau / \unit{\second}}$ \cite{icon24}.
PTB-Sr4 uses a transportable, \qty{20}{\centi\meter}--long ULE reference cavity with crystalline mirror coatings, which operates at the clock transition's subharmonic wavelength near \qty{1397}{\nano\meter} and achieves a frequency instability of \num{1.6e-16} \cite{her22};
the clock had previously reached a frequency instability $\sigma_y \approx \num{5e-16} / \sqrt{\tau / \unit{\second}}$ \cite{nos25} in local comparisons at PTB.
While the laser is frequency-doubled for interrogation of the atomic transition, its beat note with the frequency comb is measured at \qty{1397}{\nano\meter} in this work.
Spectral purity was transferred from the PTB-Sr4 cavity to the IT-Yb1 clock laser using an optical frequency comb and a digital implemntation on a field-programmable gate array \cite{sav25}, allowing IT-Yb1 to reach a similar instability.
LNEOP-SrB uses a room-temperature ULE cavity with a length of \qty{10}{\centi\metre} and a fractional frequency instability of \num{7e-16} \cite{mil09}, and reaches a fractional frequency instability $\sigma_y \approx \num{8e-16} / \sqrt{\tau / \unit{\second}}$.
The RIKEN Sr clock used spectral purity transfer from the NPL ultrastable resonator and thus achieved a similar instability as NPL-Sr1.

\subsection*{Selection of clock comparisons and baselines}
We consider two clock comparison campaigns performed for this work in 2025 as well as the data from the previous large-scale comparison of European optical clocks in 2023 \cite{icon24, piz26}, where we disregard comparisons with instabilities above $1 \times 10^{-15} / \sqrt{\tau / \si{\second}}$.
The remaining data set consists of clock comparisons PTB--NPL/LNE-OP/INRiM and a comparison NPL--LNE-OP, which we do not consider further due to its much shorter baseline.
The two comparisons highlighted in the Main Text (i.e., of NPL-Sr1 at NPL and PTB-Sr4 deployed at INRiM to the PTB-Sr3 at PTB) offer the shortest time to detection among these, whereas the other comparisons are impacted by higher clock noise or changes in the operating conditions of the optical clock network.
We further split the 2023 comparison NPL--PTB into two separate parts due to an interruption in the comparison and a change in operating conditions that resulted in a substantially higher instability of the second part (23--27 March 2023).

\subsection*{Data recording and analysis}
The recording, exchange and analysis of data follows the formalism established by Lodewyck \textit{et al.} \cite{lod20}, which represents the optical network as a graph of oscillators (e.g., clocks, transfer lasers) connected by pair-wise comparisons and thus facilitates computing the frequency ratios of clocks.
The data of the optical clocks, frequency combs and fibre links are recorded or provided at \qty{1}{\second} intervals, synchronised to local realizations of Coordinated Universal Time (UTC).
We compute the frequency ratios of the clocks during each interval and use the valid data points as input to the analysis described in the Main Text and below.

\subsection*{Stability analysis}
We estimate the white frequency noise--limited instability of an individual clock comparison by fitting the corresponding power law ($\sigma_y(\tau) \propto \tau^{-1/2}$) to the overlapping Allan deviation of reduced frequency ratio's residuals, i.e., after subtracting the RRS variations predicted by the tide model as a background signal.
The confidence intervals of the Allan deviation are calculated using the method developed in \cite{gre04}.
The fit is restricted to intermediate averaging times ($\qty{256}{\second} \le \tau \le \qty{4096}{\second}$) as the Allan deviation is governed by phase noise from fibre links and servo action in the clocks at short averaging times and may exhibit non--white frequency noise due to varying systematic biases at longer averaging times.
The Allan deviations and best fits for the two main comparisons shown in Fig.\ \ref{fig:freq-diff} as well as the five secondary ones in Extended Data Figure \ref{fig:ext:additional_clocks} are shown in Extended Data Figures \ref{fig:ext:adev} and \ref{fig:ext:additional_adevs}, respectively, while the calculated instabilities are listed in Extended Data Table \ref{tab:ext:instabilities}.

To cross-check these results, we further estimate the white frequency noise amplitude independently from the distribution of the peak amplitudes in the periodogram for frequencies well above the tidal oscillations but below the range in which phase noise becomes relevant ($\qty{40}{\micro\hertz} \le f \le \qty{1}{\milli\hertz}$).
By converting the results from PSD to power-law instability \cite{rub23}, we find similar, if moderately lower instabilities using this method.
These values are also listed in Extended Data Table \ref{tab:ext:instabilities} for all measurements.
Therefore, we conservatively use the instabilities derived from the Allan deviations, which are given in the Main Text, to compute the percentiles of the expected white frequency noise shown in Fig.\ \ref{fig:freq-diff} and discussed in the Main Text.

\subsection*{Lomb--Scargle spectral analysis}
The Lomb--Scargle periodograms used for the spectral analysis (see Figs.\ \ref{fig:freq-diff}c-d) are computed using the implementation from the AstroPy library \cite{ast22}.
We use the $\chi^2$-based
periodogram fitting algorithm without normalization in order to preserve the relative spectral powers between datasets.
The periodograms thus estimate the two-sided PSD of the signals.

We estimate the significance of peaks in the periodograms by comparing their amplitudes to the distribution of peak amplitudes that is expected for white frequency noise at the level derived for the individual comparison and the same temporal distribution.

\subsection*{Kalman filtering}
The analysis based on Kalman filtering of the measured frequency ratios provides estimates of the relative amplitude $A$ of the tidal gravity potential perturbations (see Figs.\ \ref{fig:freq-diff}e-f) as well as the fractional frequency difference $y_0$ of the clocks without these perturbations, including estimated uncertainties.

For the analysis, we first aggregate each dataset from \qty{1}{\second} to \qty{10}{\minute} averaging time by computing the mean values of non-overlapping consecutive intervals.
The averaging largely suppresses phase noise from the fibre link and leaves white frequency noise from the clocks the dominant noise contribution.
Consequently, we estimate the statistical variance of the frequency ratio in each interval from the observed overall instability and the number of valid data within the interval.
Intervals with less than \qty{30}{\percent} coverage are discarded.
The measurements $z$ and estimated variances $r$ are then processed by the filter.

We use a linear Kalman filter with the following parameters:
The system parameter $\mathbf{x} = \begin{pmatrix}y_0 & A\end{pmatrix}^\mathrm{T}$ is treated as independent of time except for process noise of the frequency difference $y_0$.
Therefore, the covariance matrix of the process noise is given by
\begin{equation}
    \mathbf{Q} =
        \begin{pmatrix}
            \sigma^2 & 0 \\
                 0   & 0 
        \end{pmatrix}
    \mbox{,}
\end{equation}
where $\sigma$ is the noise amplitude, while the state-transition matrix $\mathbf{F} = \mathbf{I}_2$ is the identity matrix.
We conservatively choose $\sigma$ such that it corresponds to a standard deviation of $5\times10^{-18}$ after 6 hours to allow slow drift of systematic effects, noting that the process noise corresponds to random-walk frequency noise.
The observation model is time-dependent and given by
\begin{equation}
    \mathbf{H} =
        \begin{pmatrix}
            1 & y_\mathrm{t}
        \end{pmatrix}
    \mbox{,}
\end{equation}
where $y_\mathrm{t}$ is the instantaneous shift as predicted by the tide model.
These are compared to the observations $z$, where the covariance of the observation noise is the statistical variance $r$ estimated during pre-processing.
The implementation of the filter is based on the \emph{filterpy} package \cite{lab15}.
We derive an initial estimate of $y_0$ by taking the mean value of the first six data points, i.e., about the first hour of valid data.
The variance of this estimate is derived from the statistical variances.
The amplitude factor $A$ is initialized as \num{0(2)}.
The Kalman filter is then fed the measurements and their estimated variances one by one and produces an updated estimate of the state and its covariance after every data point.

As discussed, we either apply the Kalman filter to the entire dataset at once or split the dataset by day and apply the filter to each day separately.
We note that, while we only report the final estimates in the Main Text, the filter outputs a progressively improving estimate with every data point.
The procedure described here can be applied to live measurement data as is.

In order to demonstrate that Kalman filter will produce null results from real data as well, we apply it to observations with negligible tidal RRS variations while providing the tidal RRS variations for a different, much longer baseline as the expected signal.
Here, we use the comparison of the clocks IT-Yb1 and PTB-Sr4 clock at INRiM in spring 2025 as measurement data.
The results are shown in Extended Data Figure \ref{fig:ext:sr4-yb-kalman}.
Due to the proximity of the clocks, the tidal RRS variations are approximately at the level of \num{e-22}.
As the expected signal, we use the predicted tidal RRS variations between either INRiM or NPL and PTB.
This results in estimates of $A = 0.41(27)$ and $A = 0.28(40)$, which are compatible with zero within about $1.5\sigma$ and $1\sigma$, respectively.

%
\backmatter

%
\bmhead{Acknowledgements}
%
We thank
    Chun Yu Ma and Mona Kempkes for support operating the ultrastable \qty{1.5}{\micro\meter} laser oscillator at PTB;
    and
    Yeshpal Singh and Kai Bongs for initiating the 2023 campaign.

\section*{Declarations}
%
\begin{itemize}
    \item \textbf{Funding}\\
    This work has been supported by
    the projects 22IEM01 TOCK,  
                 23FUN02 CoCoRICO and 
                 25IEM09 Arcsecond 
        which have received funding from the European Partnership on Metrology, co-financed from the European Union’s Horizon Europe Research and Innovation Programme and by the Participating States;
    the projects 18SIB05 ROCIT and 
                 18SIB06 TiFOON,   
        which have received funding from the EMPIR programme co-financed by the Participating States and from the European Union’s Horizon 2020 research and innovation programme;
    the Deutsche Forschungsgemeinschaft (DFG, German Research Foundation) under
        Germany’s Excellence Strategy -- EXC-2123 and EXC-2123/2 QuantumFrontiers -- Project-ID 390837967, 
        SFB~1227 DQ-\textit{mat} -- Project-ID 274200144 -- within project B02 and 
        Research Unit FOR5456 -- Project-ID 490990195; 
    the project QVLS-Q1, 
        which has received funding from the Quantum Valley Lower Saxony (QVLS), which is funded by the Volkswagenstiftung and the Ministry of Science and Culture of Lower Saxony;
    the UK Engineering and Physical Sciences Research Council (EPSRC) Grant Ref: EP/W003279/1
        for the International Clock and Oscillator Networking – ICON; 
    the UK Department for Science, Innovation and Technology
        through the National Measurement System programme; 
    the Japan Science and Technology Agency (JST) Mirai Program Grant No.\ JPMJMI18A1; 
    and
    the Max Planck--RIKEN--PTB Center for Time, Constants and Fundamental Symmetries. 
    Optical fibre links from the French REFIMEVE network are supported by Program ``Investissements d'Avenir'' launched by the French Government and implemented by Agence Nationale de la Recherche with references ANR-11-EQPX-0039 (Equipex REFIMEVE+) and ANR-21-ESRE-0029 (ESR/Equipex+T-REFIMEVE). 
    INRIM acknowledges Italian Quantum Backbone funding from the Italian Ministry of University and Research. 
    \item \textbf{Competing interests}\\
    The authors declare no competing interests.
    \item \textbf{Ethics approval and consent to participate}\\
    Not applicable
    \item \textbf{Consent for publication}\\
    Not applicable
    \item \textbf{Data availability}\\
    The data supporting the findings of this study are available within the paper, its Supplementary Information files, and the PTB Open Access Repository [DOI to be inserted].
    \item \textbf{Materials availability}\\
    Not applicable
    \item \textbf{Code availability }\\
    The code used to analyze the data in this study is available from the corresponding authors upon reasonable request.
    \item \textbf{Author contribution}\\
    The optical clocks were operated by
    \ChenHaoFeng, \MatthewJohnson and \IanHill (NPL-Sr1);
    \KilianStahl, \JoshuaKlose\ and \SoerenDoerscher (PTB-Sr3);
    \TimLuecke, \ChetanVishwakarma\ and \IngoNosske (PTB-Sr4);
    \JeromeLodewyck and \RodolpheLeTargat(LNEOP-SrB);
    \StefanoCondio, \IreneGoti, \TommasoPetrucciani and \MarcoPizzocaro (IT-Yb1);
    as well as
    \IchiroUshijima, \MasaoTakamoto, and \YogeshwarKale (RIKEN clock).
    The ultrastable lasers were operated by
    \AdamParke and \MarcoSchioppo (NPL);
    \ThomasLegero\ and \UweSterr (PTB);
    as well as
    \TimLuecke (PTB transportable laser at INRiM).
    The frequency combs were operated by 
    \XiZhang and \JacobTunesi (NPL);
    \ErikBenkler (PTB);
    \CeciliaClivati (INRiM);
    as well as
    \BenjaminPointard and \RodolpheLeTargat (LNE-OP).
    Optical fibre links were operated by
    \BenjaminPointard, \RodolpheLeTargat and \PaulEricPottie (LNE-OP);
    \BiplabDutta, \EtienneCantin, \OlivierLopez, \AnneAmyKlein, \PaulEricPottie and \ChristianChardonnet (LPL);
    \JochenKronjaeger, \AlexanderKuhl\ and \ShamboMukherjee (PTB);
    \WeiHuang and \JacquesOlivierGaudron (NPL);
    as well as
    \SimoneDonadello and \CeciliaClivati (INRiM).
    \MarcoPizzocaro, \FilippoLevi and \DavideCalonico supervised and coordinated the research at their respective institutes.
    \HeinerDenker computed the tidal variations of the gravity potential at the stations.
    \KilianStahl and \SoerenDoerscher
        performed the data analysis, using data provided by all authors.
    \KilianStahl and \SoerenDoerscher, with input from all authors, wrote the manuscript.
    All authors contributed to the discussion and interpretation of the results.
\end{itemize}

\input references.bbl

\section*{Extended Data Figures}
\begin{extfigure}
    \centering
    \rotatebox{90}{
        \includegraphics[width=\linewidth]{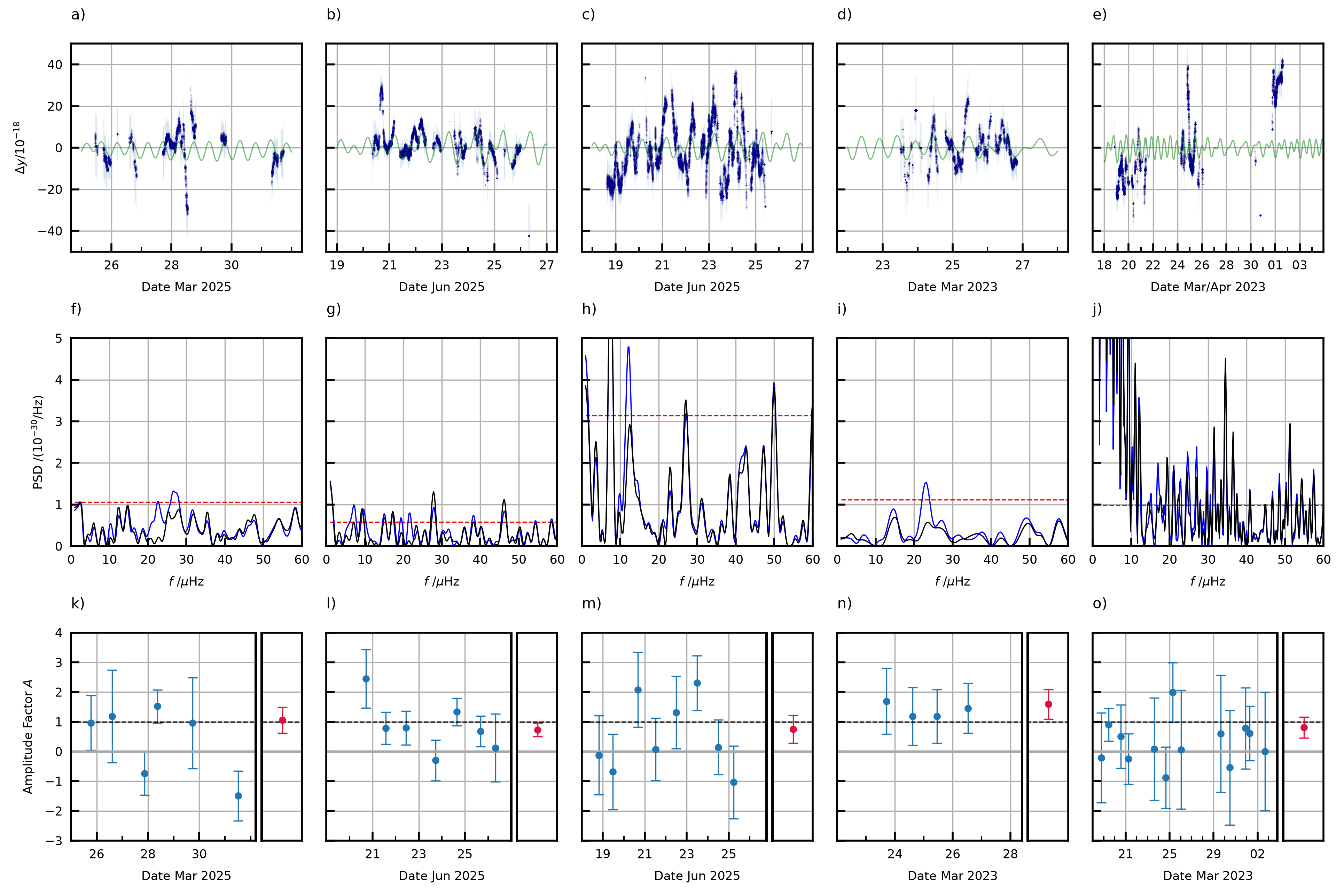}}
        \caption{
            Results of other clock comparisons (from left to right):
            IT-Yb1--PTB-Sr3 in spring 2025,
            NPL-Sr1--PTB-Sr3 in summer 2025,
            LNEOP-SrB--PTB-Sr3 in summer 2025,
            the second interval of the NPL-Sr1--PTB-Sr3 measurement reported in the Main Text
            and
            RIKEN--PTB-Sr3 where the RIKEN transportable clock was deployed at NPL.\\
            \textbf{(a-e)}
            Observed frequency ratio (blue, \qty{3}{\hour} moving average) and predicted tidal RRS variations (green).\\
            \textbf{(f-j)}
            Lomb--Scargle periodograms of the observed frequency ratio (blue) and its residuals (black), along with 99\textsuperscript{th} percentile for the estimated white frequency noise.\\
            \textbf{(k-o)}
            Amplitude estimates of the Kalman filter for individual days (blue) and the full data set (red) in the separate field to the right of each plot.
        }
    \label{fig:ext:additional_clocks}
\end{extfigure}
\begin{extfigure}
    \centering
    \includegraphics[width=1.0\linewidth]{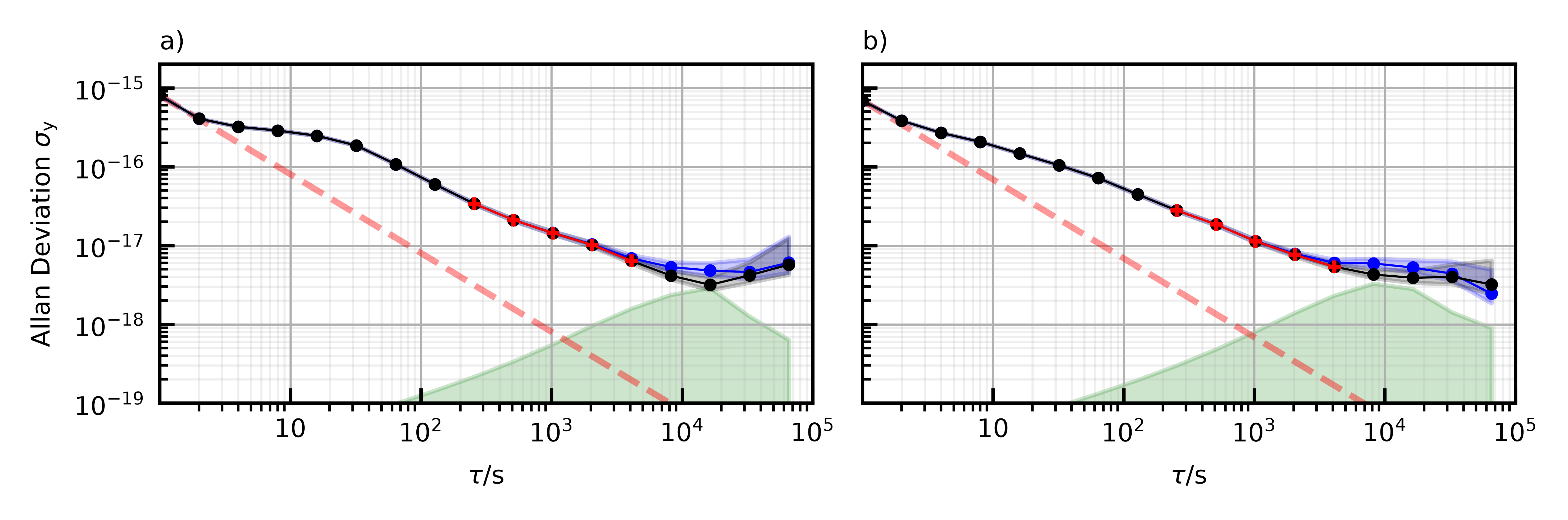}
    \caption{
        Instabilities of the the observed frequency ratios (residuals) are shown in blue (black) for \textbf{(a)} the INRiM--PTB and \textbf{(b)} the NPL--PTB  clock comparisons. 
        Shaded regions indicate the $1\sigma$ confidence intervals of the computed Allan deviations.
        The data points used to determine the white frequency noise level are highlighted in red.
        The instability caused by tidal RRS variations is shown in green. 
        The dashed red lines illustrate the estimated phase noise limit, e.g., by the fibre link.
    }
    \label{fig:ext:adev}
\end{extfigure}
\begin{extfigure}
    \centering
    \includegraphics[width=1\linewidth]{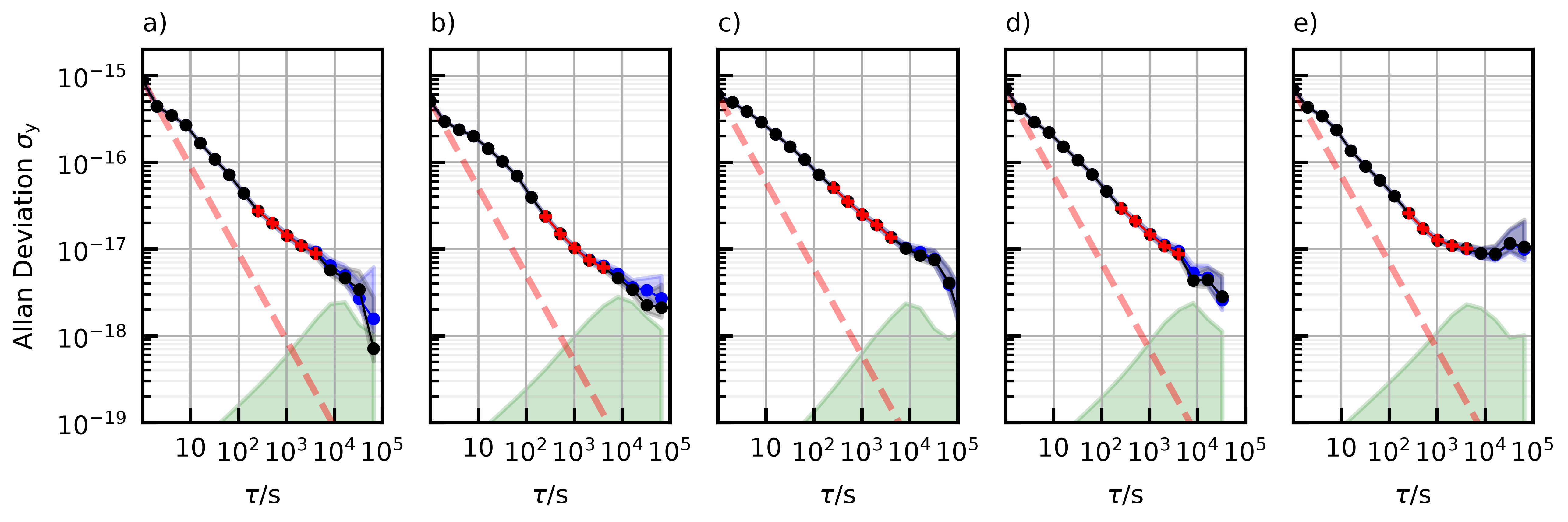}
    \caption{
        Allan Deviations of other clock comparisons:
        \textbf{(a)} IT-Yb1--PTB-Sr3 in March 2025,
        \textbf{(b)} NPL-Sr1--PTB-Sr3 in June 2025,
        \textbf{(c)} LNEOP-SrB--PTB-Sr3 in June 2025 and
        \textbf{(d)} NPL-Sr1--PTB-Sr3 in March 2023.
        \textbf{(e)} RIKEN--PTB-Sr3 in March 2023.
        See Extended Data Figure~\ref{fig:ext:adev} for further explanation.
    }
    \label{fig:ext:additional_adevs}
\end{extfigure}

\begin{exttable}[]
    \centering
    \begin{tabular}{rl|c|c}
        \multicolumn{2}{c|}{Measurement} & \multicolumn{2}{c}{$\sigma_y(\tau) \sqrt{\tau/\si{\second}} / 10^{-18}$}\\[0.5em]
        & & (ADEV) & (LS spectrum)\\
        \hline
         PTB-Sr4   & (2025)    & 467 & 407 \\
         NPL-Sr1   & (2023 I)  & 380 & 382 \\
         NPL-Sr1   & (2023 II) & 493 & 454 \\
         IT-Yb1    & (2025)    & 479 & 452 \\
         NPL-Sr1   & (2025)    & 353 & 317 \\
         LNEOP-SrB & (2025)    & 826 & 825 \\ 
         RIKEN     & (2023)    & 462 & 420
    \end{tabular}
    \caption{Estimated white frequency noise--limited instabilities of the measurements as calculated from the overlapping Allan deviation and the Lomb-Scargle periodogram peak distribution.}
    \label{tab:ext:instabilities}
\end{exttable}

\begin{extfigure}
    \centering
    \includegraphics[width=1\linewidth]{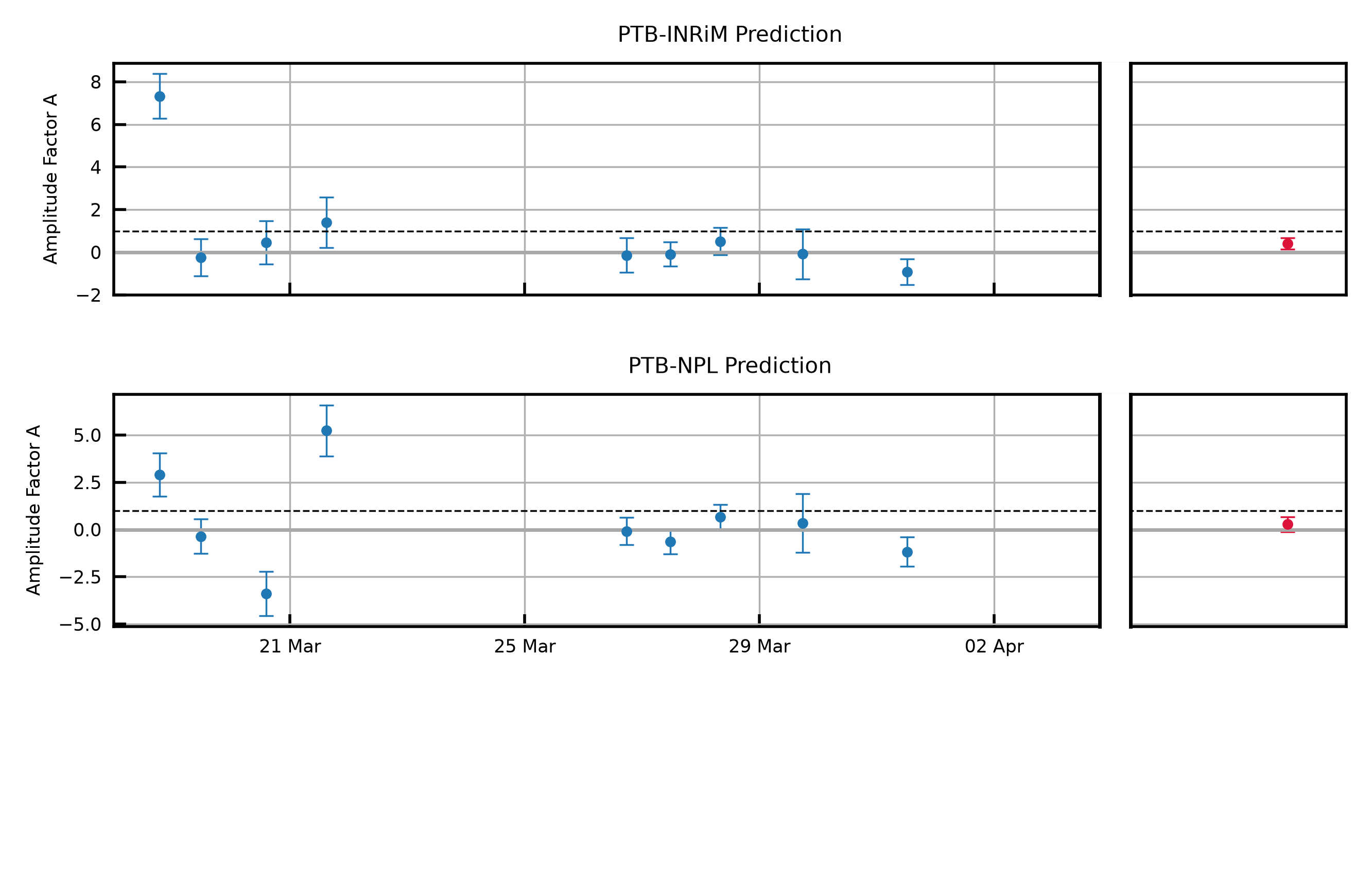}
    \caption{Null test of the Kalman filter analysis. 
    The filter was applied to the comparison of the co-located clocks PTB-Sr4 and IT-Yb1 at INRiM using the predicted tide signals for the INRiM--PTB (top) and NPL--PTB (bottom) baselines.}
    \label{fig:ext:sr4-yb-kalman}
\end{extfigure}

\end{document}